\documentclass[aps,prd,nofootinbib,twocolumn,longbibliography]{revtex4-1}
\usepackage[english]{babel}
\usepackage[autostyle, english = british]{csquotes}
\usepackage[CJKbookmarks, pdftex, bookmarksnumbered, bookmarksopen, colorlinks, citecolor=blue, linkcolor=blue]{hyperref}
\usepackage{amsmath,amssymb}
\usepackage{graphicx}
\usepackage{enumitem}
\setlist[enumerate]{topsep=0pt,parsep=-1mm,leftmargin=5mm,}
 \MakeOuterQuote{"}
\def\be{\begin{equation}}
\def\ee{\end{equation}}

\begin{document}

\title{\Large Can Alice do science and have friends, in a relational quantum world? \\
Solipsism and Relational Quantum Mechanics }

\author{Carlo Rovelli}
\affiliation{Aix-Marseille University, Universit\'e de Toulon, CPT-CNRS, F-13288 Marseille, France.}
\affiliation{Department of Philosophy and the Rotman Institute of Philosophy, 1151 Richmond St.~N London  N6A5B7, Canada}\affiliation{Perimeter Institute, 31 Caroline Street N, Waterloo ON, N2L2Y5, Canada,}
\affiliation{Santa Fe Institute, 1399 Hyde Park Road, Santa Fe, NM87501, USA.}

\begin{abstract} I discuss recent claims according to which relational understandings of quantum physics would undermine the credibility of science. 

\end{abstract}

% \pacs{04.60.Pp}
% Loop Quantum Gravity

\maketitle

1.

Some authors have stated that relational understandings of quantum phenomena [Kochen 1985, Bene-Dieks 2002, Berkovitz-Hemmo 2006, Conroy 2012, Mermin-Fuchs-Schack 2014, Healey 2012, Brukner 2015, Auff\`eves-Grangier, 2016 Zwirn, 2016, Cavalcanti 2021] and in particular Relational Quantum Mechanics (RQM) [Rovelli 1996], lead to a form of solipsism [Pienar 2021], and worried that this solipsism could undermine the possibility of doing science. 

The argument is that such solipsism appears to imply that a rational agent cannot rely on the observations of other agents. [Lewis 2004] for instance writes: ``...a theory that blocks an agent, in principle, from knowing anything beyond their momentary experiences is solipsistic in a broader, epistemic, sense. It is also rather self-undermining: the theory says that a given agent could never know whether the other physical systems presupposed by the theory behave as the theory predicts. However we categorize it, RQM is clearly deeply and problematically skeptical.'' With stronger words, [Adlam 2022] writes: ``the resulting skepticism is epistemically devastating, taking away the reasons one might have for believing in RQM in the first place.''  Even the very possibility of communicating between friends is in question, in a world described by RQM.  

The allegation against RQM, and similar views, is not to be threatened by the classical solipsism thesis. The classical question of solipsism is whether I am alone in the universe [Avramide 2023].  RQM is grounded on the opposite metaphysical assumption:  the hypothesis that I, and the experiences I have, are just one example of many other entities like me, having experiences like me. Thus RQM is quintessentially anti-solipsistic.  Rather, the allegation against RQM is that of a (somehow mis-named) epistemic form of solipsism. That is, critics claim that if RQM is true, then a cognitive agent cannot rely on the reports of other cognitive agents in order to know the facts of nature that she does not experience directly, and this undermines the possibility of doing science, or even communicating altogether. 

In this note, I inquire whether or not this line of reasoning represents a problem for RQM. The specific question I address is whether RQM undermines an agent's possibility of doing science collectively with other agents, learning, and relying upon, science. I argue here that this is definitely not the case, hence these worries are unfounded.  In a relational quantum world, agents like us can do science, do it collectively, and be friends. (See also [Cavalcanti 2021].) \\

2. 

The argument for claiming that RQM undermines the possibility of doing science can be detailed as follows. 

In the framework of RQM, a cognitive agent --let us call her Alice-- has direct access to values of physical variables ``relative to herself'', but has no direct access to values of physical variables ``relative to other physical systems''.  In particular, Alice has no direct access to values of physical variables relative to other cognitive agents --say Bob.  Alice has no direct access to Bob's experiences. 

To be sure, Alice has access to reports that she hear from Bob, about values of variables relative to him. These reports are accessible for Alice, because they are physical interactions between Bob and Alice, which in RQM are assumed to generate facts and experiences for Alice.  However, any such report is a fact relative to Alice, not a fact relative to Bob.  The two are conceptually distinct notions, and it is important to keep the distinction in mind in order to avoid apparent paradoxes raised by quantum phenomena.  Because of the discrepancy between these two notions, Bob and Alice allegedly never ``really'' communicate, and in particular  Alice cannot reliably assume that Bob's report are ``truly'' reports of facts relative to Bob.   To do science collectively, it is said, a community of observers need to have access to a shared collection of facts. These facts need to be objective, not observer dependent, that is, they must be ``absolute'', and accessible at least in principle by every observer. The experience that one observer (Bob) has about these fact must be fully reliably communicable, or accessible, to another observer (Alice).  On the other hand, the very existence of an ``absolute'' collection of facts is denied by RQM, because RQM assumes that any contingent fact about the world is only meaningful if understood as relative to some system. Therefore, it is concluded, RQM undermines the minimal requirement for doing science.   

The reason this argument is wrong is that it refers to conditions that are sufficient for doing science, but not necessary.  Science can be done successfully and collectively also under more relaxed metaphysical assumptions than the ones listed above. 

Communication between agents  has been and is a necessary ingredient for the development of the collective enterprise we call science (and for far more than that, for that matter, including having friends).  The relevant question, however, is what do we mean here by `communication'. I will argue that communication, properly understood, namely the form of communication needed to do science (and have friends), is possible in a RQM world.  

Consider the following three points.  First, according to RQM, the reports that Alice hears from Bob are informative about the world.  In fact, according to RQM, if Alice interrogates Bob and hears that that he has just measured a certain value for a certain physical quantity, then Alice can reliably expect that (unless something else has intervened in between) she will measure the same value for that quantity.  This fact is not questioned by the alleged solipsism (and does not require the additional `cross perceptive link' introduced in [Adlam-Rovelli 2023]) because it is a relation between facts relative to Alice alone: Bob's report (relative to Alice) is informative about the value (relative to Alice) of a system Bob has previously interacted with. Therefore there is a precise sense in which Bob's reports carry reliable information about the world: Alice can rely upon them.  In this precise sense, what Alice hears from other agents gives Alice a world picture which is reliable for her, according to RQM.

Second, and more importantly, imagine that Alice, ideally, does some repeated experiments and finds out a certain specific regularity in the facts relative to herself. She can then communicate to Bob that she has found this regularity.  Nothing in RQM questions this possibility.  In this case, she is not communicating a contingent fact about the world (these are relational).  She is communicating a regularity,   Regularities are assumed to be absolute aspects of reality, in RQM, not contingent and relative.  

In fact, sometimes RQM is blamed for not taking anything absolute, unlikely --say-- Special relativity, where velocity is relative by relative velocity is absolute.  This is wrong: contingent facts are taken to be relative in RQM, but their relations, mathematically expressed by the transition amplitudes of quantum theory, are not relative: they are absolute.  For instance: to have spin up in the z direction is relative, but the probability of having spin up along a certain direction if the spin was up in the z direction, \emph{this} is not relative.     And \emph{this} can be learned by Alice and communicated to Bob. 

Now, if Alice trusts RQM, she expects that Bob should find the same regularity as her.   This is because if Alice trusts RQM, she accepts the hypothesis that other agents are like her. This is the founding assumption of RQM: other quantum systems can also be observers.  So, she expects that Bob could equally find regularities and communicate the corresponding laws to her.  She has no reasons to question this, or the regularities she hears from Bob, because these are not contingent, they are not relative to Bob.  So, she could learn from Bob not only reliable facts about the world relative to her, but also, reliably, laws of nature. 

Third, the historical community of agents that has actually developed science, here on Earth, is formed by macroscopic agents that share a common decoherence domain and that store their shared information into records and memories that are described by variables that have uncertainties much larger than those questioned by the Heisenberg relations, and which therefore de facto commute.  For these reasons, the vast ensemble of facts that are accessible to the agents of this community are stable (in the sense of [Di Biagio-Rovelli 2021]) and shared. RQM does not question this fact either.

These three observations show that Alice can reliably learn (i) how the world is relative to herself and (ii) how it works in general, both from her colleagues and (iii) from past scientists. She may check by herself, if she so wishes, and, according to RQM she will confirm what she has learned.   In particular, Alice will find that the laws of nature that she has learned from Bob and in books (for instance QM laws) are reliable.  Contrary to what critics of RQM say, she has no reason to question them. 

What she cannot deduce from all that, on the other hand, is that the description of the actual state of affairs of the world that she may share with her fellow agents in the same decoherence domain, is absolute.  She cannot do so, because there could be a super-observer for whom this entire state of affairs could be just a branch in the quantum state relative to him.  That is, we can all be collectively Friend's of a Wigner. But this does not prevent us from discovering laws of nature, including quantum mechanics.\\

3. 

The subtle point in the above rebuttal of the argument purportedly claiming that RQM undermines science is the meaning of ``communicating''.   We can distinguish two separate meanings:

(i) By communicating1 we mean that we have direct access to facts relative to others. 

(ii)  By communicating2 we mean whatever we actually do when in our common language we say we are communicating among us. 

The argument according to which RQM undermines the possibility of doing science is based on the idea that to do science requires (i), while (ii), whatever it is, is insufficient.  The confusion belongs to a more general kind of mistakes, which I illustrate with an example [Rovelli 2022]. 

Say we wake up in the morning after sunrise. Sunrise is a condition for the morning to arrive and for us to wake up. For a sunrise to happen, the Sun, of course, must rise. But we learned from Copernicus that the Sun does not move, hence it cannot ``really'' rise.  We might conclude from this that there is no sunrise, therefore no morning, and therefore we should not wake up. 

That would obviously be silly.  

The reasonable thing to do is {\em to realize that what we used to call ``sunrise'' is a phenomenon that still exists, also after Copernicus. Only, it is not an actual movement of the Sun by itself, as we thought.  It is a complicated relational dance between the Sun and the spinning Earth.  Whatever it is, it does what sunrise is supposed to do: mark the time of the day for us to wake up.}

That is:  By sunrise1 we mean the sun actually moving upwards absolutely. 
By sunrise2 we mean the phenomenology that in our common language we call be sunrise, whatever it is. 

This example points to the mistake made in saying that RQM undermines science. Suppose we say: we do science by measuring observables and telling each other the results. Measuring observables and telling each other results is a condition for doing science.  The observable we measure and tell each other about, must be absolute state of affairs about reality.  But we learn from QM that there is no absolute state of affairs. We conclude that we cannot do science.  We are making the same mistake as thinking that since the Sun does not move we cannot get up in the morning.

The reasonable thing to do is {\em to realize that what we call ``measuring facts, tell each other about them'' in everyday language is a phenomenon that is still going on, also after QM.  Only, it is not an actual access to an absolute state of affairs. It is a complicated relational dance between us and the systems observed.  Whatever it is, it does precisely what measuring observables and telling each other about results has always done: be the basis for collectively doing science.}

RQM does not undermine our possibility of doing science, like Copernicus did not undermine the possibility of waking up after sunrise. 

More in general, the right question is not to ask a priori what are the conditions to do science. Or what ``real'' and ``objective'' need a priori to mean.  The right question is how we concretely did and do science.  Or what is the role that concepts like ``real'' and ``objective'' play in doing it  [Price 2024].  

Science is a collective enterprise that has happened.  At the light of certain stage of our knowledge of the world, we may have understood it in some terms (as we understood sunrise as an actual movement of the sun). But at the better light of improved knowledge we can understand it better.  For a strict follower or Aristotle and Ptolemy, the science of Galileo was not science, because what is objective are the motion of the celestial objects, and if motion is relative, these objective facts become ill defined. For a strict follower of Newton, Maxwell electrodynamics is not science, because what is objective are forces between bodies (not `fields').  As we learned from Ernst Mach, what matters are not the metaphysical prejudices we set ahead of discoveries. What matters is how we have been doing and we do science concretely, how we have concretely employed and we employ notions like ``real'' and ``objective'', and the role that these have played and play in our making sense of the world. These evolve.

This pragmatist perspective dissolves the problems raised by the authors cited at the opening of this note. This perspective is more enlightening than charging notions with a priori metaphysical weight, and then clashing with them when new knowledge arises, which does not fit into the metaphysics. Concepts should adapt to knowledge, not viceversa [Stein 2010].\\

4. 

Let us return to the critical statements presented in the first lines of this note. [Lewis 2004] writes: `` ...a theory that blocks an agent, in principle, from knowing anything beyond their momentary experiences is solipsistic in a broader, epistemic, sense.''  So far, this is just a definition of what Lewis intends for `solipsistic in a broad epistemic sense'. Then he continues: ``It is also rather self-undermining: the theory says that a given agent could never know whether the other physical systems presupposed by the theory behave as the theory predicts.''  This is factually incorrect.  Alice can predict how a system behaves with respect to her. She can prepare it, let it evolve and predict its behaviour, which she can measure.  

But she can also predict how systems behave with respect to Bob and test Bob, confirming her prediction. For instance, Alice can observe that Bob is in a $|ready>$ state.  Then she can prepare a system, send it to Bob and --knowing RQM and evolution laws implying that Bob gets entangled wit the system,-- she can predict that Bob will measure one variable or another with a certain probability distribution. From this, she can predict that if she asks (measure) Bob, she will get this or that answer with this or that probability. This is precisely what we commonly use the phrase `Alice can predict what Bob observes'. To require that what we mean is more than this is to require something that cannot be tested. 

If it cannot be tested, what is it about? 

A healthy empiricism should tell us that questions about what cannot be tested may well be addressed by recognizing that they may be meaningless.  We do not need to fatten up our ontology with untestable assertion.  

Suppose that, in the simplest situation, Alice knows (from previous measurements) the state of Bob and a system (relative to her) and (from her knowledge of the dynamics) she knows that Bob measures the system. Suppose that she then immediately makes the same measurement on the system. In a normally laboratory situation we would describe this as ``Alice has learned what Bob has measured''.   Now let us analyze the meaning of the sentence ``Alice has learned what Bob has measured''. For this sentence to be empirically meaningful, it must be empirically verifiable.  We certainly have ways of verifying the truth of this sentence if we interpret it as meaning precisely what these verification procedures commonly are, in a laboratory (asking Alice and Bob, for instance).

But suppose that, instead, we refer the fact that in RQM variables are relative and rather ask whether Alice has `really' seen the same value as Bob. Can we test this other sentence? I do not know anyway to test it empirically.   

Hence, I am happy to say it is meaningless, and discard it in the same way in learning physics I have discarded questions like:
--  Is this object `really' moving, irrespectively of references?
Or:
--  Are these two distant events `really' simultaneous, irrespectively of references?
Good physicists have learned that these questions are genuinely meaningless. In the same sense, I think that quantum physics is the discovery that whether Alice has `really' seen the same value as Bob, in an empirically untestable sense, is meaningless.  

Thinkers unhappy with this stance have deep metaphysical commitment to strong versions of realism.  I do not questioning these deep metaphysical commitments that people may have. Some people do find the lack of absolute facts in RQM incompatible with their prejudicial metaphysics.  After all, there are still people today that resist the idea that there is no non-relative meaning in ``moving'', or in ``being simultaneous''. 

My response to this reluctance is an advice to read again Ernst Mach: the success of a particular historical scientific theory should not be taken as an argument for the metaphysical necessity of its conceptual structure.  Mach's observation was liberating for Einstein and Heisenberg, both under his direct philosophical influence. Mach message is simple and effective: give up some metaphysical prejudice, and you might be able to understand the world better.  But this is up to taste and scientific nose.  Some, even today, prefer non-Lorentz invariant theories. 

What I \emph{am} questioning, on the other hand, is the argument that without such a strong versions of realism, science is impossible.   This I claim is definitely wrong.  Lewis writes: ``However we categorize it, RQM is clearly deeply and problematically skeptical.'' [Adlam] writes: ``The resulting skepticism is epistemically devastating, taking away the reasons one might have for believing in RQM in the first place.''  In light of the above, it seems clear that such jumps from the factual observation that there may be subtle quantum effects affecting reports, to the bombastic conclusion that this undermines science, are definitely too fast. 

To say that strong versions of realisms are necessary for science is like saying that unless you believe specifically in Allah you cannot be a good man (something I have often heard while traveling in Islamic countries).  A solid metaphysical foundation for one, may not be necessary for another, to obtain the same result.  We can be good men also without believing in Allah.  We can do science also without believing in absolute facts. 

I do understand that some people can hold a strong intuitive realistic metaphysics, and I respect it. Their metaphysics makes them see any worldview that renounces the heavy realist foundation of an accessible absolute state of the world as ``unconceivable''.  But what is ``unconceivable'' for one is not so for somebody else.  

To claim that a certain intuitive metaphysics is necessary for doing science has no ground.

Void of this a priori metaphysical objection, the claims that RQM undermines science are empty, at a close examination.  We can interpret QM relationally and still do science, rely on records, simply being aware that quantum effects can trick them and that there might in principle always be other branches with respect to other observers.\\

5.

The weaker form of realism required by RQM is a form of physical perspectivalism [Rodriguez 2024, Calosi-Riedel 2024].  All statements about the contingent state of reality (not their relations, the laws, the transition amplitudes) are relative to a physical system.  They concretely reflect the information available in principle to this system. A system may also have information about the information that another system has.   

Here ``information'' is taken in its physical meaning as defined by Shannon (under the name of `mutual information'): a system A has information about B iff by measuring A I can infer something about the state of B.  The statement ``A has information about B'' is itself to be understood only as relative to a system and not as absolute, namely not as necessarily valid with respect to all systems.  The statement ``the system A has information about B'' can itself be true with respect to a system (for instance, A itself) but not true with respect to another. 

In [Adlam-Rovelli 2022], a `cross-perspective' link was studied, explicitly connecting values of variables relative to Bob with Bob's report relative to Alice, partially as a reply to the objection mentioned above. 

There are two manners of understanding such `cross-perspective' link: an absolute manner, which has been particularly defended by Adlam and is sometimes denoted RQM+, and a weaker --still-relative-- manner.   The first is a genuine addition to RQM: a postulate added to the theory.   The second is only as a mere explanation, within traditional RQM, of what we mean when we say that two agents communicate. 

The absolute (RQM+) manner relies on the idea that the happening of a relative quantum event can be taken to be an absolute fact.  That is, whether an event relative to a system is realized or not is an absolute fact. This is like saying that in classical mechanics the velocity of a body A is always understood as relative to another body  B, but the fact that a body A has a velocity v relative to B is an absolute fact: it cannot be true for one observer and not true for another. This is a variation of the original form of RQM and a partial abandonment of its radical perspectivism.   RQM+ may provide a coherent picture, and it may be interesting by itself, but is not the one I have discussed here.  Here, I have focused on the relative, or weak reading of the cross-perspective link. 

The relative manner of reading the `cross-perspective' link is to understand it as a definition  within any given perspective.  This definition is the one we implicitly use when we talk about communication between agents. 

That is: if Alice assumes that the (weak) cross perspective links holds, she does not find any contradiction in talking about facts relative to somebody else.  See [Cavalcanti-Di Biagio-Rovelli 2023], where this point was also made. Using it, Alice can thus safely talk about her knowledge of values of variables relative to Bob.  

In this weaker interpretation, the link is not a modification of the perspectivalism of original RQM.  Rather, it is an observation about the extent to which Alice can make assumptions about other observer's observations, without incurring in contradiction, as we usually do in life, but always within her perspective. Namely, the link not in any `absolute' sense, independent from any reference. The absence of any meaning of the notion of absolute occurrence of a physical event was the foundational intuition underpinning the original form of RQM.

This point has been discussed in depth in [Riedel 2024, 2024b], which argues that (original) RQM leads to a perspectivalism that is not avoided by simply making quantum events relative and assuming that the occurrence of a relative quantum event is an absolute fact.  

The simple way of seeing this is to notice that, whatever we observe, we could always be in the situation of Wigner's Friend: we could always be in one of the branches of the state relative to some Wigner.  Relative to some observer, our measurement could have happened `in some branch and not in another' of our quantum state relative to him.  Hence there cannot be an absolute fact of the matter about the occurrence of a relative event. 

A hard-core realist can still postulate that all relative events are absolute facts.  It suffices to carefully distinguish absolute facts from the relative facts that can be used as inputs for the transition amplitudes giving the probabilities or events relative to a single system.  The  price to pay for holding this hard-core realism is that absolute reality is going to strongly violate intuitive aspects of our common-sense spacetime reality [Bong et al 2020].  The advantage of this position, which is dear to many, is not clear to me: what is the point of postulating something that is in principle inaccessible and is there only to reproduce something that played a role in an earlier stage of lesser knowledge?

Riedel argues that such perspectivalism may lead to a sort of infinity regress. But it is so  only if we anxiously search for an ultimate perspective.  To understand the world, ourselves, our science, and the others, we do not need such ultimate perspective.  We can simply be happy of always being, talking and describing the state of the world from within a perspective. There is no arrogance of having a privileged view here, precisely because common sense always tells us that our perspective (on contingent facts, not about laws of nature) is one among many similar others.  This is true regarding motion versus stasis, it can also be true regarding values of physical variable. 

The non-perspectival, objective information about the world contained in quantum theory is in its transition amplitudes---like in classical mechanics is its laws. Both transition amplitudes and laws are modal: they have the form ``if A then B''. They both make sense irrespectively on whether A and B are relative or absolute facts. 

Here, I have argued that even in the context of this radical perspectivalism, agents assuming RQM can do science, collect and exchange information about world's regularities, tell each other useful and effective information about the regularities of the world and also on what is to be found in the world.  

And, precisely because they know `what is like' to be one of them, they can happily communicate, and be friends.\\

**
\vskip2mm

Thanks to Eric Cavalcanti, Emily Adlam for extended discussions on the issues discussed in this note. To Huw Price and Claudio Calosi for useful exchanges and in particular to Andrea Di Biagio and Tim Riedel for very insightful reactions to early drafts of this note.  \\

{\bf References}

\addtolength{\parskip}{2mm}

Adlam, Emily (2022), ``Does science need intersubjectivity? The problem of confirmation in orthodox interpretations of quantum mechanics,'' Synthese 200: 522.

Adlam, Emily and Rovelli, Carlo  (2023). ``Information is Physical: Cross-Perspective Links in Relational Quantum Mechanics''. In: Philosophy of Physics 1.1. doi: 10.31389/pop.8. 

Auff\`eves, Alexia and Grangier, Philippe (2016) ``Contexts, Systems and Modali- ties: A New Ontology for Quantum Mechanics''. In: Foundations of Physics 46.2, pp. 121-137. doi: 10.1007/s10701-015-9952-z. 

Avramides, Anita (2023), ``Other Minds", The Stanford Encyclopedia of Philosophy (Winter 2023 Edition), Edward N. Zalta \& Uri Nodelman (eds.), https://plato.stanford.edu/archives/win2023/entries/other-minds/.

Bene, Gyula and Dieks, Dennis (2002) ``A perspectival version of the modal interpre- tation of quantum mechanics and the origin of macroscopic behavior''. In: Foundations of Physics 32, pp. 645-671. doi: https://doi.org/ 10.1023/a:1016014008418. 

Berkovitz, Joseph and Hemmo, Meir (2006) ``A new modal interpretation in terms of relational properties''. In: Physical Theory and its Interpretation - Essays in Honor of Jeffrey Bub. Ed. by William Demopoulous Itamar Pitowsky. Springer, pp. 1-28. doi: https://doi.org/10.1007/1-4020- 4876-9\_1. 

Bong, K.W., Utreras-Alarc\'on,  A., Ghafari, F., Liang, Y.C, Tischler, N., Cavalcanti,   E.G., Pryde, G.J., 
Wiseman  H.M.  (2020) ``A strong no-go theorem on the Wigner's friend paradox'', Nature Physics 16, 1199. ArXiv: 1907.05607

Brukner, Caslav (2015). ``On the quantum measurement problem" arXiv: 1507. 05255 [quant-ph]. 

Calosi, Claudio and Riedel, Timotheus (2024) ``Relational Quantum Mechanics at the Crossroads". Foundation of Physics 54, 74

Cavalcanti, Eric G.  (2021) ``The View from a Wigner Bubble" Foundation of Physics 51, 39. arxiv:2008.05100.

Cavalcanti,  Eric G., Di Biagio, Andrea and Rovelli, Carlo ``On the consistency of relative facts'' The European Journal for Philosophy of Science, 13 (2023) 4, 55. arXiv:2305.07343. 

Conroy, Christina (2012). ``The relative facts interpretation and Everett's note added in proof''. In: Studies in History and Philosophy of Science Part B: Studies in History and Philosophy of Modern Physics 43.2, pp. 112-120. doi: https://doi.org/10.1016/j.shpsb.2012.03. 001. url: https://www.sciencedirect.com/science/article/pii/ S1355219812000196. 

Di Biagio, Andrea and Rovelli, Carlo (2021) ``Stable Facts, Relative Facts'', Foundations of Physics, 51, 30. arXiv:2006.15543. 

Healey, Richard (2012) ``Quantum Theory: A Pragmatist Approach''. In: British Journal for the Philosophy of Science 63.4, pp. 729-771. doi: 10. 1093/bjps/axr054. 

Kochen, Simon (1985). ``A New Interpretation of Quantum Mechanics''. In: Sym- posium on the Foundations of Modern Physics 1985. Ed. by Peter Mittelstaedt and Pekka Lahti. World Scientific, 1985, pp. 151-169. 

Leifer, Matt (2018). What are Copenhagenish Interpretations and should they be Perspectival? (talk). url: www.youtube.com/watch?v=CC\_K- %20gK6q4. 

Mermin, David, Fuchs, Christopher and Schack, Rudiger (2014) ``An introduction to QBism with an application to the locality of quantum mechanics''. In: American Journal of Physics 82.8, pp. 749-754. doi: 10.1119/1.4874855.https://pubs.aip.org/aapt/ajp/ article-pdf/82/8/749/13089031/749\ \_1\ \_online.pdf. url: https: //doi.org/10.1119/1.4874855. 

Peter J. Lewis (2024) ``A dilemma for relational quantum mechanics'', Dartmouth College, August 13.

Pienaar, Jacques, (2021),``A quintet of quandaries: Five no-go theorems for relational quantum mechanics,'' Foundations of Physics 51: 97.

Price, Huw (2024).  Conference on pragmatism, London Ontario.

Riedel, Timotheus (2024) ``Relational Quantum Mechanics, quantum relativism, and the iteration of relativity'', Studies in History and Philosophy of Science 104, 109-118.

Riedel, Timotheus (2024b) ``Is Quantum Relativism Untameable? Revenge Wigner Arguments for Relative Facts''. In: British Journal for the Philosophy of Science (forthcoming). doi: 10.1086/732830. 

Rodriguez, Pascal (2024) ``Bohrian Perspectivism'', HOPOS 2024, Vienna.

Rovelli, Carlo (1996) ``Relational quantum mechanics',' International Journal of Theoretical Physics 35: 1637-1678.

Rovelli, Carlo (2022) ``The Old Fisherman's Mistake'', Metaphilosophy 53 (2022) 567-746, http://philsci-
archive.pitt.edu/18837.

Stein, H. (2010), ``How Does Physics Bear upon Metaphysics; and Why Did Plato Hold that Philosophy Cannot be Written Down?'', Colloquium Talk at the University of Chicago.  See A.W. Carus, The Pragmatics of Scientific Knowledge: Howard Stein's Reshaping of Logical Empiricism, The Monist, Vol. 93, No. 4, Philosophical History of Science, 618-639.

Zwirn, Herv\'e (2016 ) ``The Measurement Problem: Decoherence and Convivial Solipsism''. In: Foundations of Physics 46.6, pp. 635-667. doi: 10. 1007/s10701-016-9999-5.

%\section{Introduction}

\bibliographystyle{/Users/carlo/Dropbox/utcaps}
\bibliography{/Users/carlo/Dropbox/library.bib}

\end{document}